\documentclass[]{aastex62} 
\usepackage{amsmath,amstext}

\newcommand{\okina}{\textquoteleft}

\newcommand{\Ou}{{\okina}Oumuamua}
\newcommand{\be}{\begin{eqnarray}}
\newcommand{\ee}{\end{eqnarray}}
\usepackage{xcolor}
\usepackage{comment}
\graphicspath{{./}{figures/}}

\received{February 4, 2019}
\revised{March 11, 2019}
\accepted{\today}
\submitjournal{ApJL}

\shorttitle{\Ou s Make Planets}
\shortauthors{Pfalzner \& Bannister}

\begin{document}

\title{A hypothesis for the rapid formation of planets}

\author[0000-0002-5003-4714]{Susanne Pfalzner} 
\affiliation{J\"ulich Supercomputing Center, Forschungszentrum J\"ulich, 52428 J\"ulich, Germany}
\affiliation{Max-Planck-Institut f\"ur Radioastronomie, Auf dem H\"ugel 69, 53121 Bonn, Germany}

\author[0000-0003-3257-4490]{Michele T. Bannister}
\affiliation{Astrophysics Research Centre, Queen's University Belfast, Belfast BT7 1NN, United Kingdom}

\correspondingauthor{Susanne Pfalzner and Michele Bannister}
\email{s.pfalzner@fz-juelich.de, michele.t.bannister@gmail.com}

\begin{abstract}
The discovery of 1I/\Ou\ confirmed that planetesimals must exist in great numbers in interstellar space. Originally generated during planet formation, they are scattered from their original systems and subsequently drift through interstellar space. As a consequence they should seed molecular clouds with at least hundred-metre-scale objects. We consider how the galactic background density of planetesimals, enriched from successive generations of star and system formation, can be incorporated into forming stellar systems. We find that at minimum of the order of 10$^{7}$ \Ou-sized and larger objects, plausibly including hundred-kilometre-scale objects, should be present in protoplanetary disks. At such initial sizes, the growth process of these seed planetesimals in the initial gas- and dust-rich protoplanetary disks is likely to be substantially accelerated. This could resolve the tension between accretionary timescales and the observed youth of fully-fledged planetary systems. Our results strongly advocate that the population of interstellar planetesimals should be taken into account in future studies of planet formation. As not only the Galaxy's stellar metallicity increased over time but also the density of interstellar objects, we hypothesize that this enriched seeding accelerates and enhances planetary formation after the first couple of generations of planetary systems. 
\end{abstract}


\section{Introduction} 
\label{sec:intro}
The detection of 1I/2017 U1 \Ou\ \citep{Meech:2017} confirmed the long-standing hypothesis that interstellar objects (ISOs)\footnote{Here the term `ISOs' describes free-floating planetesimals with a size $\gtrsim 50$~m.} 
should be common in the interstellar medium (ISM) \citep{Whipple:1975, Sekanina:1976,Torbett:1986,McGlynn:1989,Stern:1990,Kresak:1992, Jewitt:2003,Francis:2005, MoroMartin:2009, Cook:2016,Engelhardt:2017}, since a vast portion of every planetary system's small bodies are ejected or gently drift away over the lifetime of a star. At the extreme end this also includes free-floating planets; though at an estimated occurrence rate of 0.0096--0.18 per star \citep{Fujii:2018}, they are rare compared with \Ou-sized objects. 
The ISM and molecular clouds are therefore enriched by past generations of planetesimals --- as much a part of this process as the well-studied elemental enrichment by supernovae.
The small mass fraction of heavy elements in planetesimals even merits consideration in how they affect the chemical evolution of the Galaxy \citep{Tinsley:1974}.

\begin{figure*}[ht]
\plotone{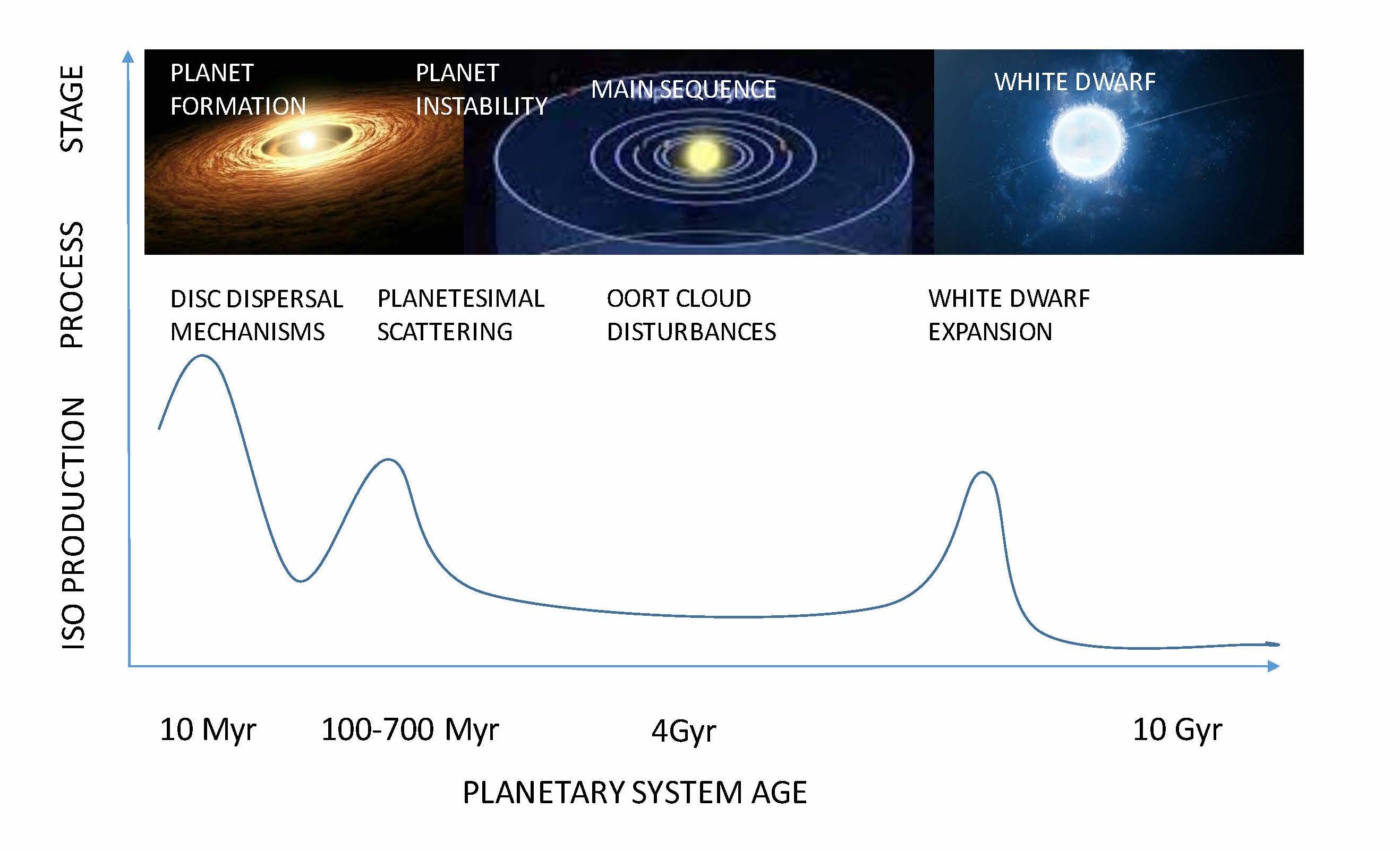}
\caption{Schematic picture of interstellar object (ISO) production rate, relative to the population of the initial planetesimal disk, during the lifetime of a solar-type star. For high-mass stars the last stage of enhanced planetesimal formation would happen much earlier at the end of the lifetime of the star.
}
\label{fig:ISO_production}
\end{figure*}

It seems that planetary systems shed their planetesimals to interstellar space as inevitably as dandelions cast their seeds on the wind (Fig.~\ref{fig:ISO_production}).
Early on, predominantly icy planetesimals are liberated from the outer disk by interactions with neighbouring stars, due to the high stellar density during that phase \citep{Pfalzner:2015,Hands:2019}. As the gas disk dissipates and the system's planets migrate, close encounters with giant planets lead to the ejection of a large portion of a system's remaining planetesimals \citep{Duncan:1987,Charnoz:2003,Raymond:2017}. These planetesimals reside between a few au to tens of au from their star and, therefore, will be a mix of rocky and icy bodies. Afterwards, throughout the star's lifetime, predominately icy planetesimals (rocky objects like those observed by \citet{Meech:2016} are 4\% of the Solar Oort cloud; \citealt{Shannon:2015}) will be gently lost, as they drift from the distant fringes of the star's Oort cloud under the nudging of the Galactic tide and passing field stars \citep{Brasser:2010, Kaib:2011,Hanse:2018}. Finally, the remainder of the system's Oort cloud will be shed to interstellar space once the star leaves the main sequence and loses mass \citep{Veras:2011,Veras:2014,Do:2018, MoroMartin:2019}. 
All these ISOs wander the Galaxy, at velocities ranging from very near their local standard of rest (LSR) up to the cutoff of Galactic escape velocity, and it appears the great majority are retained for Gyr \citep{Guilbert:2015}. 
Multiplied across the successive generations of stellar system formation, wandering ISOs form a steadily increasing Galactic background. 

In the context of planetary system formation, so far attention has focused on exchanging planetesimals \emph{after} system formation \citep[e.g.][]{Adams:2005,Levison:2010,Belbruno:2012,Jilkova:2016}. As the birth material of circumstellar disks is inevitably described as ``dust particles of at most a micrometer in size" \citep{Birnstiel:2016}, the effect of the pervasive population of ISOs on the formation of planetary systems has received scant consideration so far. \citet{Grishin:2018} considered the capture of ISOs into already-formed gas-rich disks.
However, stars and their surrounding planetary systems form from molecular clouds and ultimately the ISM.
Here we consider how the presence of ISOs in the source ISM and molecular clouds, through cluster development into fully-fledged systems, eventually affects planet formation around the stars.

\section{The current picture of planet formation} 
\label{sec:planetform}

Two main pathways for the formation of planets have been extensively discussed \citep[for reviews, see, for example][]{Blum:2008,Morbi:2016,Armitage:2018}.
In the first scenario, planets form when spiral arms become gravitationally unstable and fragment directly to form large protoplanets within just a few thousand to ten thousand years \citep{Boss:2001,Kratter:2016}. However, for typical protoplanetary disks, the material near the star stays too hot to go unstable, so this process is expected to generate planets typically only at large distances $\geq$ 100~au from the star \citep{Kratter:2016} with masses in the range of 10--20 Jupiter masses.
Most planets detected so far orbit their host star at distances of $<10$~au, and have considerably smaller masses. Even taking migration in the disk into account, gravitational instability is unlikely to be the predominant formation process for the majority of planets. 

For planets that reside close to their star, dust accretion is the standard formation scenario \citep{Armitage:2018}. 
Here microscopic dust grains grow by sticking collisions first into larger porous aggregates, later cm-sized ``pebbles", and eventually planetesimals, from which terrestrial planets or the cores of gas giants form \citep[e.g.][]{Youdin:2015}.
Two potential problems with this scenario have been found: i) a relatively long formation timescale, especially for giant planets, and ii) growth barriers during the accretion process. 
The accretion model requires timescales of $10^2-10^4$~yr to form mm- to cm-sized pebbles, $10^4-10^6$~yr until the planetesimal stage is reached, $10^6-10^7$~yr to form terrestrial-type planets, and an additional $10^5$~yr for the gas giants to accumulate their gas \citep{Pollack:1996,Armitage:2018}. 
This seems at odds with observations of the disk frequency in young clusters, which indicate the median protoplanetary disk lifetime to be merely 1--3 Myr for both dust and gas \citep{Haisch:2001,Mamajek:2009,Richert:19}. However, there also exist counterarguments; individual disks have order-of-magnitude scatter from 1--10~Myr, and this derived timescale might be biased by selection effects \citep{Pfalzner:2014}.
Nevertheless, planets have been confirmed in $\sim$~Myr-old gas-rich disks \citep{Johns-Krull:2016,Keppler:2018} 
and at $\sim$~Myr-old disk-less pre-main-sequence stars \citep[e.g. V830 Tau b;][]{Donati:2016}.
Fast planet formation is also favoured by interpretations of the ring and gap structures commonly observed in very young ($<1$~Myr; perhaps as little as 100 kyr) disks like HL Tau in high-resolution ALMA observations \citep{Andrews:2018} as carved by (proto)planets \citep[e.g.][]{Long:2018, Zhang:2018, vanderMarel:2019}. 
At least some planets must form faster than in the originally proposed planetesimal accretion model. 

The second, but connected, problem concerns the existence of several growth barriers --- the bouncing, fragmentation and drift barriers --- in the original accretion model \citep{Weidenschilling:1980,Brauer:2007,Blum:2008,Zsom:2010}. The bouncing and fragmentation barriers occur for sizes $<0.1$~m, while modelling of the later stages require bodies of at least 100~m in size to form planets \citep{Fortier:2013}. However, the streaming instability \citep{Youdin:2005,Simon:2017} and pebble accretion \citep{Lambrechts:2012,Kretke:2014,Levison:2015a,Levison:2015b,Johansen:2018} are currently regarded as promising ways to overcome these barriers \citep{Armitage:2018}.
The streaming instability describes a linear instability in aerodynamically coupled mixtures of particles and gas that leads to small-scale clustering of the solids, while pebble accretion is a complementary process where the strong gravitational focusing of pebbles onto larger embryos drives their rapid growth.

The starting point of the various accretion models is dust grains of  $\approx 0.1-1.0$~\micron, growing into mm-sized particles, as inferred from measurements at infrared wavelengths of interstellar extinction \citep{Testi:2014}. However, as only particles with sizes comparable to the observational wavelength are actually detectable, ISOs are invisible in these measurements. Given the evidence for rapid planet formation, we consider if it is plausible that ISOs act as nucleation centres to jump-start accretion in protoplanetary disks.

\section{ISO abundance from the interstellar medium to the individual star}
\label{sec:ISO_star}

\begin{figure*}[t]
\plotone{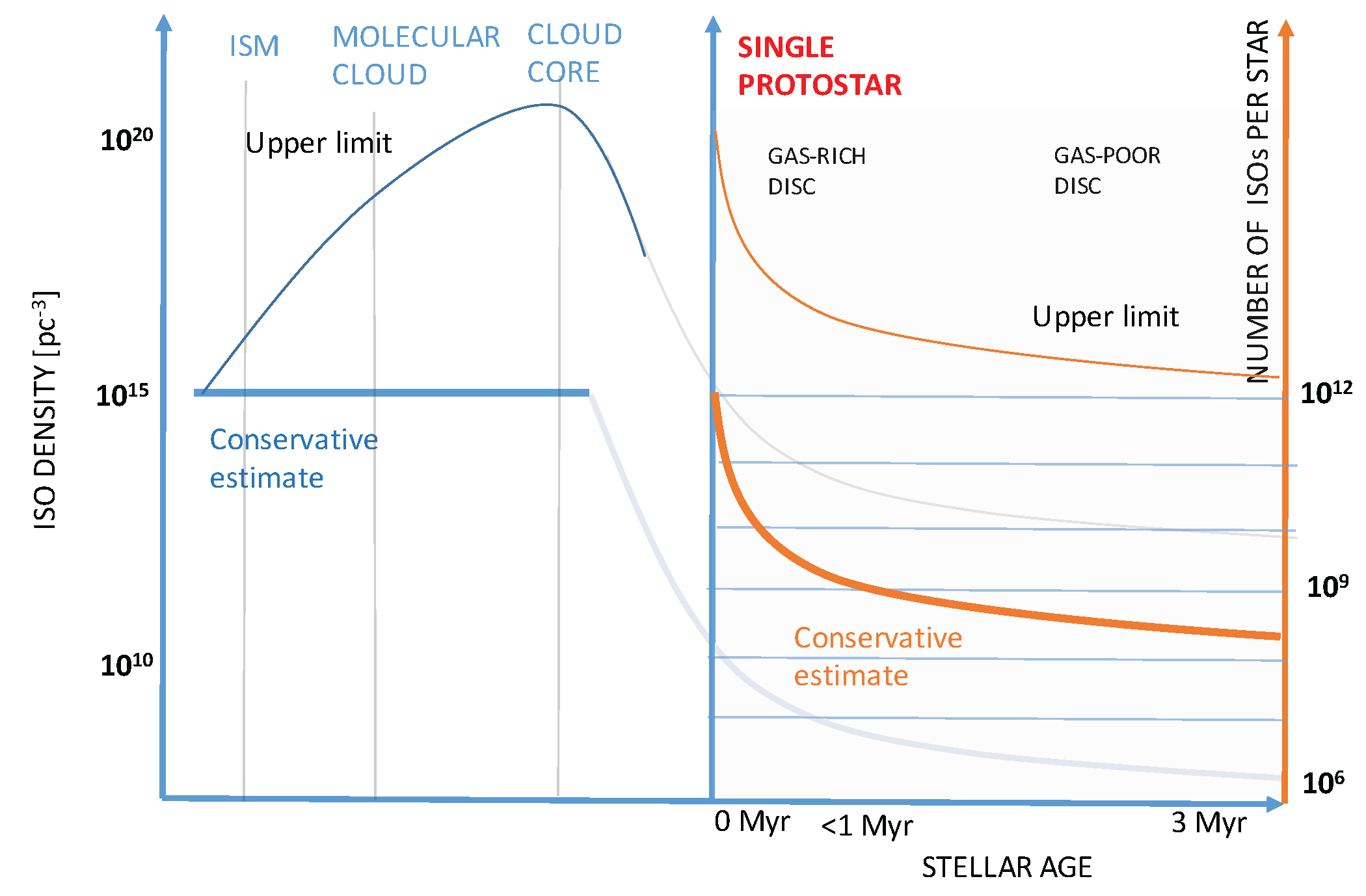}
\caption{Illustration of the seeding of planetesimals from the interstellar field density into disks. ISO density indicates $\sim 100$~m-sized effectively-inactive objects, like \Ou.
}
\label{fig:planet_form}
\end{figure*}

We start by considering the ISOs available in an arbitrary cubic parsec of the ISM, and follow the development all the way to protoplanetary disks, as outlined in Fig.~\ref{fig:planet_form}. We adopt a conservative approach in the sense that we take the lowest possible values at each stage. We emphasize that even if only a relatively small number of ISOs are incorporated early on in the disk, these will suffice to substantially accelerate planet formation.

{\bf Initial density, option (i): ISM}
ISOs float effectively invisible to us through the ISM, only detectable in the event of a chance encounter with the Sun as in the case of \Ou. 
Estimating their occurrence rate is challenging, requiring complex assessment of surveys' detection efficiency for both active (brighter) and inactive ISOs. 
The most comprehensive available observational constraint for the local spatial field density of \Ou-like ($\sim100$~m inactive\footnote{\Ou's size is constrained by the {\it Spitzer} non-detection at an effective radius of 49--220 m, including an effective radius of 70 m for a surface with cometary scattering properties \citep{Trilling:2018}. \Ou's visual brightness was not appreciably enhanced by its weak outgassing inferred by \citet{Micheli:2018}.}) ISOs is $\rho_{\rm{ISO}} \simeq 10^{15}$ pc$^{-3}$ (\citealt{Meech:2017} from 30 integrated years of Pan-STARRS1/Catalina Sky Survey observation, and scaling from sizes of 1 km to 100 m, after \citet{Engelhardt:2017}; note that PS1 discovered \Ou).
This is compatible with the {\it active} limit from LINEAR comets of \citet{Francis:2005}. 

For simplicity we assume that this ISM density is representative for the Galaxy, fully aware that it probably varies throughout the Galaxy; with higher values in the Galactic bulge, disk and towards the Galactic center due to more frequent dynamical encounters \citep{Veras:2014}, more metal-rich Pop I stars \citep{Tanikawa:2018} and gravitational focusing effects.  
In addition, impact erosion by interstellar dust grains might generate some localized fluctuations in the ISO population by removing very small ISOs over the course of several tens of Myr \citep{SternShull1990, Stern:1990, Vavilov:2019}, which is compatible with the deficit of very small comets in the size distribution of long-period comets \citep{Meech:2004}. In contrast, $\gtrsim10$~km-sized ISOs will be long-lived, effectively for the Galactic lifetime, as they experience only a minute reduction in their size \citep{Guilbert:2015}. 
An additional loss process may be gas-flow-induced torques in the ISM, which could spin-up some ISOs to disruption \citep{Hoang:2018}; finally, dynamical capture by fully evolved planetary systems removes only a negligible fraction of the population \citep{Torbett:1986,Engelhardt:2017}.
However, there is no reason based on the Sun's current location for the ISO density in the solar neighbourhood to be locally enhanced, and the Sun has probably been at its current Galactic distance for around a Gyr. Therefore, we shall assume that our locally measured ISO number density is characteristic for the ISM, and that it is isotropic and uniform within our cubic parsec. 

{\bf Initial density, option (ii): molecular clouds}
The ISO concentration could be enhanced in molecular clouds (MCs) in comparison to the ISM as part of the MC forming process, or by capture into already-formed MCs.  As MCs form, the gas density typically increases by a factor of 10$^4$--10$^6$. The question is whether ISOs are concentrated to the same degree as the gas. MC formation has been attributed to different processes, including converging flows driven by stellar feedback or turbulence, gravitational instability, magneto-gravitational instability, and instability involving differential buoyancy (for a review, see \citet{Dobbs:2014}).
Gravitationally-driven processes act as effectively on ISOs as on the gas, but this is not the case for magnetically-driven processes. The relative importance of the different processes is still under debate.

ISOs will only be efficiently captured into molecular clouds if their velocity \emph{relative} to that of the gas is not too large.  Unfortunately, there exist a host of uncertainties regarding the \emph{relative} velocities of clouds and ISOs. Most importantly, the ISO velocity distribution is currently not constrained by observations. \Ou\ itself had a velocity very close to the local standard of rest prior to its interaction with the solar system. However, it is unclear whether \Ou's velocity is typical for ISOs, especially as \citet{Engelhardt:2017} noted that low-velocity ISOs are preferentially detectable, so a selection effect is at work.

As ISOs come from stellar systems, one would expect their velocity distribution to be similar to the stellar velocity distribution, which depends on the stellar age. As we will discuss in Sec.~\ref{sec:overtime}, it is likely that planet, and therefore planetesimal formation, became more efficient over time; as such it can be assumed that most ISOs come from stars $<$ 2 Gyr. These stars have a lower velocity dispersion than older stars, but are generally more dynamically excited than that of MCs, with rich substructure and complexities \citep[e.g.][]{Hayden:2019,Mackereth:2019}. The various ISO production processes (Fig.~\ref{fig:ISO_production}) will give them velocities relative to that of their parent star, in the range 0--10~km/s for ejection from young planetary systems \citep{Adams:2005,Hands:2019} and $<0.5$~km/s 
for the gently unbound late-stage ISOs \citep[D. Veras pers. comm. per][]{Veras:2014vel}.
The velocity distribution of each star's contribution of ISOs will then be broadened over time by encounters; for instance, \Ou's velocity has been altered by its Solar encounter from near the LSR out to a more typical dispersion from the local mean \citep{Meech:2017,Feng:2018}.
In summary, the formation mechanisms for ISOs would suggest that their velocity distribution is slightly wider than that of the stellar velocity distribution.

When considering how fast molecular clouds move relative to their nearby stars, past studies focus on the internal properties of the clouds rather than their kinematics in the disk. 
On the other hand, the internal one-dimensional velocity dispersion of molecular clouds can be approximated as $\sigma_{\rm{1D}}= 1.2 \rm{D}^{0.3}$ km $s^{-1}$ = 0.6--4.8 km $s^{-1}$, where D=0.1--100 is the typical diameter of the clouds. This agrees with the observationally measured values that lie in the range of 3--10 km $s^{-1}$ \citep{Stark:1984,Miville-Deschenes:2017}. In a simple model for the possibility of MC capture one can assume that only the ISOs with relative $v < 10 \rm{km/s}$ can be captured in a cloud. \citet{Mackereth:2019} found using multivariate Gaussian distributions that  $\sim5$\% of stars they sampled had velocities of $\lesssim 10$~km/s; these stars have ages $<2$~Gyr. Under the above assumptions, this would imply that 1-4\% of ISOs can be captured into MCs.
Thus, even without a full model of the ISO velocity distribution and its comparison to a detailed Galactic map of MC bulk velocities, it is nevertheless plausible that MCs directly capture a fraction of ISOs and as such additionally raise the ISO density. Due to the diverse uncertainties the ISO density in MCs is not well constrained, but likely is in the range of 10$^{15}$ to 10$^{19}$ pc$^{-3}$.

{\bf From the molecular cloud to the clump}
Gas density increases again by a factor of 10$^2$--10$^3$ when parts of a molecular cloud become unstable and form clumps \citep{Bergin:2007}. If ISOs are concentrated in the same way that gas is in clumps, the ISO density in clumps could increase even to $10^{22}$~pc$^{-3}$. However, here the question arises of whether only the ISOs with low relative velocities are captured. Gravitational capture during the collapse of the cloud happens if the ISOs' velocities are below the escape speed from the cloud, meaning
$    v(t) < \sqrt{2GM/d(t)}    $ , with $d$ being the clump diameter. 
Additionally, ISOs crossing in from outside our cubic parsec could contribute to the ISO content after collapse starts. The crossing time of our parsec of MC for relative velocities of $0.1-20$~km/s will range from 50 kyr to 10 Myr, less than the cloud freefall time of 10 Myr \citep{Miville-Deschenes:2017}. 
Capture effectively erases the ISOs' velocity distribution.

In summary, there is potentially substantial ISO concentration once clumps have formed, with plausible densities of up to $10^{21-22}$~pc$^{-3}$. However, there are also extreme uncertainties about the relative effects of the different processes; to be on the safe side we assume no ISO concentration at all from the ISM to the clump stage, staying at the ISM value of $\sim 10^{15}$~pc$^{-3}$.

{\bf From clumps to stars}
An entire cluster of stars usually forms near-simultaneously from such a clump, allowing us to estimate the available number of ISOs available per star.
For the solar neighbourhood, the star formation efficiency typically transforms 10\%--30\% of the clump mass into stars.
Most stars in such a cluster have a mass of 0.5 M$_\sun$, therefore to form a cluster of a hundred stars requires a clump  mass of $M_{cl}\approx$  170--500 M$_\sun$; similarly, forming a cluster of five thousand stars would require a clump of $ M_{cl}\approx $ 8500--25000 M$_\sun$. 
There is a direct correlation between the mass of a clump and its size \citep{Urquhart:2014,Pfalzner:2016}:
\be 
\log(M_{cl}) = 3.42 \pm 0.01 + (1.67 \pm 0.025) \times \log(R_{cl}).
\ee
with $M_{cl}$ in units of solar masses and $R_{cl}$ in pc. Using this correlation, one finds that the clump for a hundred stars would typically have a radius $R_{cl}(100)\approx$  \mbox{0.3--0.4 pc}, whereas those forming five thousand stars have $R_{cl}(5000) \approx 1.0-1.2$~pc. 
Thus, the volume $V$ of clump space that each individual star draws its material from is $V=4 \pi R_{cl}^3/3N$, and seems to be relatively independent of the cluster mass between 0.001--0.002 pc$^3$ at least for the dominant mode of clustered star formation.
Applying a typical star formation efficiency of 30\% and our adopted ISO density of $\rho_{ISO} \approx 10^{15}$~pc$^{-3}$, the number of ISOs available per forming star,  $N^S_{ISO}$ is therefore,
\be N^S_{ISO} = 0.3 V \rho_{ISO} \approx \mbox {3-6} \times 10^{11}.\ee
This number should be regarded as a lower limit, as the ISO density in the molecular cloud could be as much as a factor of 10$^6$ higher (as discussed earlier in \S~\ref{sec:ISO_star}), and consequently the same applies for the number of ISOs per star.

\section{The availability and effect of planetesimal seeds in the disk}
\label{sec:growth}
The next step is the formation of a protostar and its surrounding disk, where first a gas-rich disk forms from which a gas-poor disk develops. Here again the ISOs will behave similarly to the gas as long as the processes are dominated by gravitational forces, but differ if gas-dynamical or magnetic forces play an important role.  
We shall oversimplify the complex processes governing disk formation by assuming that the ISO density develops similarly to that of the gas; future investigation will have to confirm how far such an assumption is justified. 
The inclination distribution of ISOs will damp down and settle into a thin midplane layer in $\ll$Myr; more slowly than the orbital timescale, but much less than the disk lifetime.
At the end of the disk formation process, the mass of the disk is typically 0.01-0.1 times the mass of the star \citep{Andrews:2013}. If we assume that this ratio is also typical for ISOs, it means that most of the ISOs will actually end up in the star, but despite this low efficiency, the disk typically still contains $\approx 10^{8}-10^{9}$ ISOs.  
We term these `seed ISOs' to distinguish that these are embedded planetesimals.

The number of seed ISOs in the disk could be altered by several processes.
Heating could destroy some of the smaller $\lesssim 1$~km ice-rich seed ISOs, with the efficiency of this process dependant on the seed ISO size, composition, and their location in the disk (i.e. the temperature they experience). The disk area that is cold enough ($T<$ 100K) to retain water and other volatiles  depends on the mass of the host star; for solar-type stars it is outside a few au for solar-type stars \citep{Walsh:2012} and moves to $\sim20$~au for higher mass stars. In any case there are large volumes of cold disk available. Particle exchange between inner and outer areas will be quite small as turbulent mixing in T Tauri disks is minimal \citep[e.g.][]{Willacy:2015}. Therefore, we can assume that this processing leaves some larger seed ISOs with surface devolatilization, but otherwise unaffected. 
The total number of available seed ISOs could also increase during the gas-rich disk phase, due to capture of ISOs from the cluster environment.  The effectiveness of aerodynamic drag capture is weak for $0.1-10$~km-sized ISOs, but quite efficient for $\sim$10-metre ISOs, sourcing only tens of $>0.5$~km seed ISOs \citep{Grishin:2018}, which probably reshapes the size-distribution of the seed ISOs. Like the gas, the ISOs probably have a higher density in the inner disk, which is also where most potential loss processes take place.  The combined effects of heating and capture need to be modelled in detail; we infer it leads to a loss of very small seed ISOs. However, even if we assume that only 10 per cent of seed ISOs are retained during this stage, $\mathcal{O}(10^7-10^8)$ seed ISOs with sizes $\gtrsim 100$~m will still be present in the accretion stages.

The sizes of the seed ISOs embedded in the disk will control both their survival times during the disk's $10^6-10^7$~yr gas-rich phase \citep{Alexander:2014} and their efficiency of accretionary growth. 
At present, the size distribution of the entire ISO population remains unconstrained by our knowledge of only \Ou, and moderate constraints will require the discovery of tens of more ISOs by surveys such as LSST \citep{Cook:2016}. 
The cumulative size distribution will depend on the collisional evolution that each ISO has undergone, which will in turn depend on its dynamical history in its origin system.
The number density of ISOs cannot yet be reliably related to the mass-density production of planetary systems, as there remain substantial caveats concerning the choice of ISO size distribution function \citep[see detailed discussion by][]{ISSI:2019}.
The size distribution's precise form is non-critical for our hypothesis: we need only obtain a guide for the behaviour of our comparatively tiny sample in a disk.
Reasonable size distribution assumptions come from the well-constrained broken power-law distribution functions observed among the various Solar System minor planet populations from $\sim 1-1000$~km \citep[e.g. discussions in][]{MoroMartin:2009,Belbruno:2012}. 
Applying the broken power law with break radii at $r_b = 3$~km, 30~km, 90~km for the number of seed ISOs $n(r)$ of radius $r$, after \citet{MoroMartin:2009}: 
\begin{align}
\begin{split}
    n(r) &\propto r^{-q_{1}} \text{ if } r < r_{b} \\
    n(r) &\propto r^{-q_{2}} \text{ if } r > r_{b} 
\end{split}
\end{align}
with $q_1= 2.0$ to 3.5 and $q_2 = 3$ to 5, respectively, we consider geometric albedos of 0.04 (cometary) through $p_v \sim 0.1$ (\Ou; \citep{Trilling:2018}).
Across this parameter space, for our number density of $N(D > 100$~m) $\approx 10^{7}$ seed ISOs, almost all will be \Ou-sized objects, with plausibly $10^4$--$10^5$ objects with diameter $D\sim 1$~km, some $10^3$ objects with $D\sim 100$~km, and even a couple of dwarf planets per planet-forming disk.
These estimates are sufficiently robust if instead single power-law distributions are considered.

The majority of the seed ISOs are thus large enough to have long survival times in the disk. 
Hundred-metre-scale seed ISOs will have a gas-drag-induced drift timescale of $10^5$~yr in the inner part of disks at a few au, rising to $10^7$~yr by $\sim 50$~au \citep{Weidenschilling:1977}; longer than the inclination-damping timescale, and certainly long enough for accretionary growth processes. 
The thousand-km seed ISOs are up into the regime where the planet-disk interaction is dominated by gravity and Type I migration can occur \citep{Goldreich:1979}.  
However, they are sufficiently small to only experience minimal migration \citep[e.g.][]{Levison:2015a}, with a timescale of Myr, as the migration timescale decreases linearly with planet mass.
It is also plausible that various inhomogeneities of the disk will halt their inward migration \citep[e.g.][]{Cridland:2019}.
The effects of the seed ISOs on the young disk will be challenging to discern observationally. 
Single objects of this size are too low-mass to open an annular gap, nor are they directly detectable in continuum emission. They would generate low-amplitude Linblad spiral density waves, which may be detectable in optical observations of disk eclipses. 

With the seed ISOs stably located in the disk for long durations, their size distribution provides progenitors of planet(esimals) at a range of different sizes. 
At the small end of the size distribution that contains most of the seed ISOs, the provision of $\simeq 10^6$ 100-m and $10^5$ km-sized objects in the gas-rich disk may help in initially resolving the streaming instability's low efficiency at producing km-size planetesimals, due to the effects of turbulent diffusion \citep{Nesvorny:2018}.
The effect on the streaming instability of emplacing 100-m objects has not yet been modelled; this is a difficult size for resolving in simulations.
It remains to be seen if this density in the disk of small ISO seeds is sufficient for efficient collisionary accretion.  
In contrast, among the $\simeq 10^3$ planetesimals of size 100-km and larger, any ISO seeds that are $\geq 200$~km in size will start to grow toward rapid pebble accretion \citep{Visser:2016,Johansen:2017}.
Such sizes have been modelled as cores for viscously stirred pebble accretion: embedded in a gas-rich disk that slowly forms pebbles, the initially largest objects are likely to dominate, and rapidly generate terrestrial and giant planets in $10^3$ years \citep{Levison:2015a, Levison:2015b}.
Thus the largest of these nascent worlds will grow efficiently \citep{Youdin:2005,Johansen:2007, Ormel:2010,WindmarK2012a}.

Not all ISO seeds will grow into planets. 
Some may be scattered from the system after a period of accretion and return to the ISO population. 
Without sufficient collisions, the small end of the ISO seed size distribution may not grow at all.
The simulations of \citet{Levison:2015a} imply even Pluto-sized objects would survive intact without growth in the 20--30~au region of a Solar System-like disk. Thus, our hypothesis predicts the Solar System's trans-Neptunian populations could contain former ISOs --- at low probability only, considering the losses during planetary formation and migration.

\section{The temporal development of planet formation}
\label{sec:overtime}

As the overall ISO population of the Galaxy will have gradually built up over Gyr, modulo the removal of ISOs into new planetary systems as we propose here, the first generation of stars will have lacked a background of planetesimals.
This does not trouble our hypothesis: the known variety of planetesimal formation processes would take place in first-generation protoplanetary disks, simply on longer timescales than we observe in current-generation disks. This implies that in the past, planet formation would have been slower, and potentially less efficient, so the planet population and planetary system structure could have changed over time. The planetesimals would have been ejected by the processes illustrated in Fig.1 but at slightly later stellar ages.
The observed higher frequency of planets around high-metallicity stars than around low-metallicity stars \citep{Fisher:2005} and indications that planet properties depend on metallicity \citep{Narang:2018} are natural outcomes of our scenario. 
Equally, the non-detection of planets in globular clusters can be interpreted as a lower planet formation rate during that epoch, rather than the potential alternatives of a bias against detection due to the high stellar density, hindered planet formation \citep{Vincke:2018} or increased planet ejection due to the high stellar density.

\section{Conclusion}
\label{sec:conclusion}

We have shown that the estimated numbers of \Ou-like objects in the interstellar medium implies that their presence could accelerate the planet formation process considerably. Some small and icy ISOs are likely to be removed by various erosion and heating processes. 
Given the substantial uncertainties in several of the evolutionary steps, at least $\mathcal{O}(10^7)$ ISOs of hundred-metre-scale and larger size can survive into the gas-rich disk.
In particular, uncertainties in whether the formation of molecular clouds concentrates ISOs from their ISM density could mean these values are several orders of magnitude higher in the disk.
These embedded planetesimals would then function as seeds for fast and efficient planet formation.
The overall ISO population of the Galaxy will have gradually built up over billions of years. We only broadly outline this scenario, but it seems highly necessary that star and planet formation scenarios take into account the abundant presence of ISOs throughout the Galaxy, and their increase on Gyr time scales.

This seeding scenario also implies that planet formation was slower in the earliest generations of stars. Planetary differentiation will surely disperse the original ISO material, which is also $\ll0.1\%$ of the mass of even a terrestrial planet; the bulk planetary compositions will be dominated by that of the disk. 
Yet ISOs from an ancient star may have once been the hearts of many young planets.

\acknowledgments

M.T.B. acknowledges support from UK Science and Technology Facilities Council grant ST/P0003094/1. 
We appreciate helpful conversations with Sean Raymond, Richard Alexander, Bertram Bitsch, Jos de Bruijne, Alan Fitzsimmons, Paul Francis, Samantha Lawler, Ted Mackereth, Mordecai-Mark Mac Low, Tom Millar and Dimitri Veras. 
We thank the two referees for their thoughtful comments that helped improve and strengthen the manuscript.
We thank the International Space Science Institute (ISSI Bern), which made this collaboration possible, and our enthusiastic colleagues on the ISSI \Ou\ team\footnote{\url{http://www.issibern.ch/teams/1ioumuamua/}} for an enjoyable workshop that helped spark this work.

\bibliographystyle{aasjournal}
\bibliography{references}

\end{document}